# GRAVITATION AND THE PROBLEM OF QUANTUM MEASURMENT

Chris Allen Broka
2002 Granemore St. Las Vegas, NV 89135
(chris.broka@gmail.com)

## Abstract

We consider the possibility that the goal of quantizing General Relativity should be abandoned in favor of Semiclassical Gravity. A formalism is provided for doing so. This paper addresses what happens when von Neumann measurements take place in this context. Particular attention is paid to the EPR paradox and the Page-Geilker experiment.

## Introduction.

Despite much effort the quantization of General Relativity has remained an elusive goal. The non-linearity of the theory poses serious problems. And if we would try to picture gravity as mediated by spin-2 gravitons we encounter another problem—the resulting theory cannot be renormalized in any conventional sense (1, 2). String Theory gives rise to massless spin-2 particles that could be identified with gravitons. But the mathematics is difficult and the project remains a work in progress. Loop Quantum Gravity is a completely different suggestion (3).

Semiclassical Gravity is the theory that results when we treat spacetime in a classical sense but require the quantized fields to obey its geometry. It is generally required that $G_{\mu\nu} = 8\pi \langle\Psi|T_{\mu\nu}|\Psi\rangle$ where $T_{\mu\nu}$ is the operator representing the stress-energy tensor for the quantum field theory of interest and $|\Psi\rangle$ the state of its matter fields. Many interesting results have followed from this approach. These include Hawking radiation (4), and the creation of particles in an expanding universe (5). Nevertheless, it has largely been regarded a sort-of poor halfway house between classical physics and a proper theory of quantum gravity.

## Semiclassical Gravity.

Perhaps our wish to quantize gravity should be resisted. Such ideas have been discussed in the past (6, 7). This requires us to assume the existence of a classical spacetime manifold whose geometry is given. It also requires us to specify upon this a quantum field theory (QFT) we are interested in. From this we get equations of motion for the quantum fields. We solve these equations and get expressions for our quantum fields that are to be written in terms of the creation and annihilation operators appropriate to the Fock space that defines our quantum field theory. For now we work in the Heisenberg Picture. Write the action as:

1) $\int (-R / 16\pi + \mathcal{L}_{\text{field}}) \sqrt{-|g|}\, d^4\boldsymbol{x}$ from which follows:

2) $T_{\mu\nu} = 2\,\frac{\delta\mathcal{L}_{\text{field}}}{\delta g^{\mu\nu}} - g_{\mu\nu}\,\mathcal{L}_{\text{field}}.$

Equation 2) ensures that the divergence of $T_{\mu\nu}$ vanishes (8). We assume $\mathcal{L}_{\text{field}}$ has no explicit spacetime dependence.

Our Fock space must contain and be constructed around a vacuum state $|0>$. In even the simplest theories $\int <0\,\big|\,T_{00}(\boldsymbol{x}, t)\,\big|\,0> d^3\boldsymbol{V}$ diverges. (Here $d^3\boldsymbol{V}$ designates the volume element for the spacelike hypersurface of interest.) We will address this in a very simple and straightforward way:

3) $G_{\mu\nu} = 8\,\pi\,(<\Psi|T_{\mu\nu}|\Psi> - <0_M|T_{\mu\nu}|0_M>) = 8\,\pi\,\{\Psi|T_{\mu\nu}|\Psi\}$ where $\{\Psi|\mathbb{O}|\Psi\}$ is defined as $<\Psi|\mathbb{O}|\Psi> - <0_M|\mathbb{O}|0_M>$ for any operator $\mathbb{O}$. If $|\Psi> = |0_M>$ $G_{\mu\nu} = 0$ and we are in Minkowski space.

This becomes our new Einstein's equation. We might better have written the action as:

1') $$\int (-R\,/\,16\,\pi + \{\Psi\,|\,\mathcal{L}_{\text{field}}\,|\,\Psi\})\sqrt{-\,|\,g\,|}\;d^4\boldsymbol{x}.$$

## Simple Cases.

Consider a very elementary example. Spacetime is a Minkowski manifold; $g_{\mu\nu} = \eta_{\mu\nu}$. Suppose we are interested in a real scalar field $\varphi(\mathbf{x}, t)$. Suppose $\mathcal{L}_{\text{field}} = \frac{1}{2}(\eta^{\mu\nu}(\varphi_{,\mu})(\varphi_{,\nu}) - m^2\,\varphi^2)$. From this we get a Klein-Gordon equation whose solution is familiar. We may write:

4) $\varphi(\mathbf{x}, t) = \frac{1}{\sqrt{V}} \sum_k \frac{1}{\sqrt{2\,\omega_k}}(e^{-i\,k\cdot x}\,a_k + e^{i\,k\cdot x}\,a_k^\dagger)$ where $\omega_k = \sqrt{\boldsymbol{k}^2 + m^2}$ and we imagine the system contained in an enormous periodic box of volume $V$.

This allows us to construct a simple Fock space whose basis vectors consist of $|0_M>$, 1-particle states $|\mathbf{k}>$, two-particle states $|\mathbf{k}, \mathbf{k}'>$, and so on. We find that $\{\Psi|\int T_{00}(\boldsymbol{x}, t)\,d^3\boldsymbol{V}|\Psi\} = <\Psi|\sum_k\,\omega_k\,a_k^\dagger\,a_k\,|\Psi>$ which is finite for a state like $|\mathbf{k}>$. It occurs to us that we have already found a solution to our above-discussed problem. And we did not have to look very far for it. $\{\mathbf{k}|T_{\mu\nu}|\mathbf{k}\} = 0$ since Minkowski space is infinite ($V \to \infty$). Therefore $G_{\mu\nu} = 0$. This is a completely adequate situation according to our criteria as long as there are only finitely many particles represented in $|\Psi>$.

Although $|\Psi>$ is not in any way a function of spacetime it can contain information relevant to it. Consider a Klein-Gordon state— $|\Psi_{\text{KG}}> = \frac{1}{\text{N}} \sum_k \frac{e^{-i\,k\cdot x_0}}{\sqrt{2\,\omega_k}}|\mathbf{k}>$ (where N is for normalization). $\{\Psi_{\text{KG}}|\,T_{00}(\boldsymbol{x}, t)\,|\Psi_{\text{KG}}\}$ describes a world in which we have a massive particle localized at $\boldsymbol{x}_0$. This particle will curve the spacetime around it so the manifold cannot be Minkowskian. We have produced an inadequate and inconsistent situation. We would have to try other manifolds in the hope of finding one that gave us a consistent solution. Since this problem seems rather simple we think it could be solved (given a little cleverness and patience). But it is not obvious how to solve such problems, in general, other than by trial and error.

Let us consider another simple case—the Einstein-de Sitter cosmology. Here our manifold is spatially flat and has $g_{00} = 1$, $g_{\text{ii}} = -\,t^{4/3}$, with the rest 0. From this we derive a curved spacetime version of the Klein-Gordon equation whose solution we can write as:

5) $\varphi(\mathbf{x}, t) = \frac{1}{\sqrt{V_0}} \sum_k \frac{1}{t}(u_k\, a_k + u_k{}^*\, a_k^\dagger)$.

Unfortunately, we are unable to find simple solutions for $u_k$ except when k = 0. Fortunately, these are the only solutions we will end up requiring. Let us write:

5') $\varphi(\mathbf{x}, t) = \frac{1}{\sqrt{V_0\, t^2}}\, \frac{1}{\sqrt{2\, m}}\, (e^{-i\, m\, t}\, a_0 + e^{i\, m\, t}\, a_0^\dagger)$ + other terms.

$V_0\, t^2$ represents the comoving volume element for this cosmology.

From equations 2) and 5') we can deduce $T_{\mu\nu}$. We must now find a |Ψ> that affords the desired expectation values for these operators. Since the matter in this universe is distributed evenly and is at rest in our coordinate system let us guess that $|\Psi> \, = a_0^\dagger\, |0_M>$; essentially we have put a single, k = 0, massive particle *everywhere* in this universe. We posit $a_0\, |0_M> = 0$ . We find:

6) $\{\Psi|T_{00}|\Psi\} = \rho = \frac{m}{\mathrm{Vo}\, t^2} + \frac{1}{\mathrm{Vo}\, m\, t^4}$ (which contains a $1/t^4$ term). All other $\{\Psi|T_{\mu\nu}|\Psi\} = 0$.

We know from Einstein's equation that $\rho = 1/(6\, \pi\, t^2)$. So we must set $m = \mathrm{Vo}/6\, \pi$. And we recognize that Vo is infinite. The unwanted $1/t^4$ term vanishes and we are left with the result we desire. We are able to find a perfectly acceptable solution to this problem. But it does require us to adopt a rather strange QFT—one in which infinite mass particles exist.

In these examples we have worked in the Heisenberg picture. This has been practical because we have only dealt with simple Lagrangians that give rise to quantum fields that satisfy linear equations whose solutions can be interpreted easily in terms of creation and annihilation operators the physical meaning of which we can pretty well understand. But physically interesting Lagrangians contain complicated, nonlinear, interaction terms. We would have a hard time calculating useful results were we forced to work in anything besides the Dirac Interaction Picture. Let us write $H(t) = \int T_{00}(\boldsymbol{x}, t)\, d^3 \boldsymbol{V} = \int (\mathcal{H}_0\, (\boldsymbol{x}, t) + \mathcal{H}'\, (\boldsymbol{x}, t)\,)\, d^3 \boldsymbol{V} = H_0(t) + H'(t)$ where the former designates the underlying (linear) part of the QFT and the latter the interaction terms. It is from the former that we derive creation and annihilation operators that make intuitive sense to us. It is these operators for which we assume $a_k\, |0_M> = 0$. Creation operators are associated with negative-frequency solutions and annihilation operators with their positive-frequency counterparts. |Ψ> must now be written |Ψ*(t)*> where $i\, \partial_t\, |\Psi(t)> = H'(t)\, |\Psi(t)>$. The existence of interactions will, in many cases, require us to regularize and renormalize our QFT. All operators here (e.g. $T_{\mu\nu}(\mathbf{x}, t)$) must, themselves, be represented in the Interaction Picture. We also note that unitary evolution is not possible here. The curvature of spacetime will change both $\varphi(\boldsymbol{x}, t)$ and $H(t)$ in non-linear ways and Einstein's equation is non-linear as well. If |Ψ*(t)*> evolves into $|\Psi_1(t)>$ and $|\Psi_2(0)>$ into $|\Psi_2(t)>$ taking gravity into account, it will not usually be true that $\alpha\, |\Psi_1(0)> + \beta\, |\Psi_2(0)>$ evolves into $\alpha\, |\Psi_1(t)> + \beta\, |\Psi_2(t)>$.

## Discussion.

What is suggested can be cast in the form of a simple protocol:

A) Specify a globally hyperbolic spacetime manifold $\mathcal{M}^4$. It can be whatever one likes. Impose upon it a

coordinate system (*t, x, y, z*) which can be anything one likes provided *t* allows us to define a set of spacelike Cauchy hypersurfaces.

B) Define upon it a QFT as desired (i.e. define $\mathcal{L}_{\text{field}}$) and solve its field equations given $\mathcal{M}^4$ thus providing expressions for $T_{\mu\nu}$. $T_{\mu\nu}$ is given by equation 2).

C) Find a |Ψ> in the Fock space corresponding to the QFT such that $G_{\mu\nu} = 8\,\pi\,\{\Psi|T_{\mu\nu}|\Psi\}$.

One could supplement this protocol with additional requirements (e.g. the weak energy condition) if one wished.

We have already examined some simple cases where it is possible to accomplish this goal. We must, of course, hope that the actual $\mathcal{M}^4$ we live in has a QFT and |Ψ> that make all of this consistent.

## Quantum Measurement, EPR, and the Page-Geilker Experiment.

In this paper we are interested primarily in what happens when a von Neumann measurement occurs leading to what is familiarly (if a little carelessly) known as wave function collapse. In earlier work (16) this author has attempted to frame the problem in term of the Interaction Picture and the evolution of the Fock space state of our world |Ψ*(t)*>. Normally it evolves by unitary evolution — $i\,\partial_t\;|\Psi(t)> \;=\; H'(t)\;/\Psi(t)>$. But not all |Ψ*(t)*> are admissible in this interpretation. I give the example of an electron going through a Stern-Gerlach apparatus. In one scenario it comes in spin-up and strikes a detector that makes a light turn red. A different scenario has a spin-down electron triggering a green light. Had the electron started out (|+> + |->)/$\sqrt{2}$ unitary evolution would lead to a superposed state having us seeing both a green and a red light. This is considered an inadmissible state (since we cannot be conscious of the green light and the red one simultaneously) and /Ψ*(t)*> is projected into either the 'red' or 'green' state in accordance with the Born rule. It will be noted that /Ψ*(t)*> contains all the information relevant to the electron, the measuring device and the observer. It will also be noted that I invoke consciousness in distinguishing between admissible and inadmissible state vectors. So I am dealing with a variant of the von Neumann-Wigner Interpretation related somewhat to Chalmers' theory of M-Properties (9). Some readers may regard the involvement of consciousness in physics as anathematic. They can consider consciousness as a placeholder here that could signify other criteria they might prefer (e.g. complexity, the simple size of the measuring device, gravity itself (10)). In any case, there just seem to be states of reality that are not allowed. What happens when a wave function "collapses?"

In earlier work (11) we proposed that our Fock space consisted of a collection of admissible states, $\{C_i\}$, with the rest being inadmissible (such as those in which observers simultaneously witness both a green and red light). We worked in the Interaction Picture and paid no attention to General Relativity. We introduced a "projection operator," $\mathfrak{S}$, which, if |Ψ*(t)*> were to evolve into an inadmissible state, would convert it into one of the $C_i$ at random with a *relative* probability given by the absolute square of $<C_i|\Psi(t)>$. Since the time coordinate here is specific to a particular Lorentz frame we might worry that Special Relativity could suffer violation. We tried to argue that this would not be the case. But the matter was left somewhat up in the air. We can sharpen this argument by supposing that $\mathfrak{S}$ can only project |Ψ*(t)*> into such $C_i$s that:

7) $\{\Psi(t)\ \mathfrak{S}^{\dagger} \left|T_{\mu\nu}(\mathbf{x},\ t)\right|\ \mathfrak{S}\ \Psi(t)\} = \{\Psi(t)\ |T_{\mu\nu}(\mathbf{x},\ t)|\ \Psi(t)\}$ whenever $(\mathbf{x},\ t)$ lies outside the future light cone of the measurement event.

This addresses one troubling question very nicely. We might worry that we chose the wrong Lorentz frame to describe our Stern-Gerlach experiment. If we had chosen a different one $|\Psi(t)>$ would have been projected on a different spacelike hypersurface (also, of course, containing the measurement event). But, according to 7) this makes no difference. No observed physics could change anywhere outside the future light cone of the measurement event and this light cone is independent of our choice of Lorentz frame. It does not matter how we decide to look at it.

It could also be objected that our projection scenario would not, in general, give us the Born Rule. This is entirely correct. The various $C_i$s into which $|\Psi(t)>$ might find itself projected will not usually be orthogonal to one another — $|\Psi(t)>$ describes not only the Stern-Gerlach apparatus and its electron but, also, the enormous, possibly infinite, number of other particles that inhabit our universe. Our small experiment represents only a miniscule "part" of $|\Psi(t)>$ and we would expect $| < C_i \mid \Psi(t) > |^2$ to be nearly 1 irrespective of our experiment provided that projection into $C_i$ did not involve much change to the rest of the universe. But this does not reflect the way in which actual quantum measurements are performed. Suppose some experimentalist wanted to determine the energy of a superposed photon state consisting of a red photon and a green one by making a von Neumann measurement. (Red photons have less energy than green ones.) They would evacuate their apparatus, surround it with thick lead shielding, and, perhaps, place it in an abandoned salt mine. They would do everything possible to insulate it from any interaction with the rest of the universe. They try to create a sort-of "mini-universe" consisting of only their photon. To the extent they succeed, their red-photon and green-photon states really can be considered as orthogonal and the Born Rule will be recovered. If they tried to perform their experiment outside on a sunny day they would, of course, not find the Born Rule to work at all. They would probably just be observing random photons coming from the Sun.

But suppose more than one measurement event occurs. We give an example based on the Aspect experiment (12). Suppose, at the origin of their Lorentz frame, some physicists have a device that can produce a superposed and entangled photon state consisting of two green photons or two red ones. One photon moves off to the left. The other one goes to the right. Suppose a measurement event occurs at a leftward point $P_1$ and the another at a rightwards point $P_2$. Stipulation 7) cannot work here since, were the photon at $P_1$ found to be green $|\Psi(t)>$ could not have its behavior effected at $P_2$ since these events are spacelike separated. A red photon could, therefore, be detected at $P_2$ and we know from Aspect's result that such cannot be the case. We, clearly, must amend 7) under these circumstances. We will say that $\{\Psi(t)\ \mathfrak{S}^{\dagger} \left|T_{\mu\nu}(\mathbf{x},\ t)\right|\ \mathfrak{S}\ \Psi(t)\} = \{\Psi(t)\ |T_{\mu\nu}(\mathbf{x},\ t)|\ \Psi(t)\}$ whenever $(\mathbf{x},\ t)$ lies outside the causal future of the two (or more) measurement events. By 'causal future' here we mean the union of the future light cones originating from $P_1$ and $P_2$. Now the green measurement event at $P_1$ will project $|\Psi(t)>$ into a $C_i$ such that a green result must be obtained at $P_2$.

We mention, in passing, that another way can be found to overcome our difficulty (13, 14). We could say that, when two measurement events take place, $|\Psi(t)>$ projects so as to allow physics to be changed within their common causal past. In our example we would say that $|\Psi(t)>$ projected at the origin where the entangled state was first prepared. For this to happen the universe must "know" that the two measurements are foreordained to take place and effect must precede cause. We much prefer our answer. But retrocausality was, in fact, proposed as a resolution of the EPR paradox immediately after the appearance of EPR's paper. We would wonder why $\mathfrak{S}$ would project the state at the origin since there nothing obviously inadmissible about it there.

This may explain the red and green lights and photons. But what is the case with gravity? An experiment of Page and Geilker (15) looked for an answer. They, essentially, did as above with the lights replaced by a gravity detector and a heavy lead sphere that could be moved to either point A or B depending on the outcome of a quantum measurement. A |+> electron would move it to A and a |-> one would result in its going to B. (This is a sort-of simplification of the actual experiment.) Would a (|+> + |->)/$\sqrt{2}$ electron result in a situation where gravity would be coming from A and B simultaneously? No. They find that this is not the case. Just as with the lights, only one of the two "classical" possibilities is observed. They interpret this as suggesting that gravity itself needs to be quantized. This is one way of looking at it, but not the only one. Spacetime in their laboratory was very close to Minkowskian since the sphere was not sufficiently massive to curve it very much. Something very like unitary evolution would, therefore, hold between the measurement of the electron spin and the measurement of the sphere's gravity. Perhaps the experimenters could simply not perceive a ghostly half-sphere at both A and B. And they had to determine the gravitational field with some kind of gravitometer. Say that, if the sphere was at A an indicator would point to -1. If it was at B the pointer would go to 1. They could not very well see a ghostly half-pointer pointing at both 1 and -1. It may have been to avoid such "absurd" situations that $\mathfrak{S}$ projected the state vector as observed. We could also appeal to Instrumentalism. Maybe the large and complex device that measured the electron and moved the sphere could, itself, not exist in a superposed state.

## Conclusion.

Part B) of our "Protocol" requires that we find solutions to our field equation that work within our $\mathcal{M}^4$. This may or may not be possible. Very importantly, we require a Fock space. This, in turn, requires that $[a_k, a_{k'}] = [a_k^\dagger, a_{k'}^\dagger] = 0$ and $[a_k, a_{k'}^\dagger] = \delta_{kk'}$ for our creation and annihilation operators. If we posit as much we will not, in general, recover the canonical equal-time commutation relations we might want. We could say 'too bad for the equal-time commutation relations.' But this might lead to other consequences (e.g. the failure of microcausality). And our protocol leaves us with a mathematically near-intractable problem — how to satisfy it. We have given a few examples where we satisfied it by, essentially, lucky guessing. But, faced with a truly difficult and interesting problem (such as the one posed by the very complicated $\mathcal{M}^4$ we really live in), lucky guessing is not likely to help much.

In this paper we have tried to address the problem of von Neumann measurements as they would occur in $\mathcal{M}^4$. We think the idea expressed in 7) is a good one and likely to hold regardless of some of the problematic issues mentioned above. And we extend it to situations where more than one measurement is made on the system. The result may look, to some readers, very much like 'spooky action at a distance' (which, in fact, it is). But, if we are to accept the Aspect results, this may just be what we have to settle for.

## Acknowledgement.

The author is grateful to Professor W. G. Unruh (UBC) for interesting and helpful comments regarding this work.

## References.